\documentclass[epjST]{svjour}
\usepackage{graphicx}
\usepackage{bm}
\usepackage{amsmath}
\usepackage{amssymb}

\begin{document}

\title{Time-dependent Ginzburg-Landau model for light-induced superconductivity in the cuprate LESCO}

\author{M. Ross Tagaras, Jian Weng, and Roland E. Allen}

\institute{Department of Physics and Astronomy, Texas A\&M University, College Station,
TX 77843}

\abstract{Cavalleri and coworkers have discovered evidence
of light-induced superconductivity and related phenomena 
in several different materials. Here we suggest that some features may be
naturally interpreted using a time-dependent Ginzburg-Landau model. In
particular, we focus on the lifetime of the transient state in
La$_{1.675}$Eu$_{0.2}$Sr$_{0.125}$CuO$_4$ (LESCO$_{1/8}$), 
which is remarkably long below about 25 K, but exhibits different behavior 
at higher temperature.}

\maketitle

\section{Introduction}

In this brief note we suggest that time-dependent Ginzburg-Landau models may
be useful in interpreting the experiments of Cavalleri and coworkers (and
other groups) that have demonstrated ultrafast phase transitions in
materials responding to femtosecond-scale laser pulses.

\begin{figure}[tbp]
\centering
\includegraphics[width = 0.49\columnwidth]{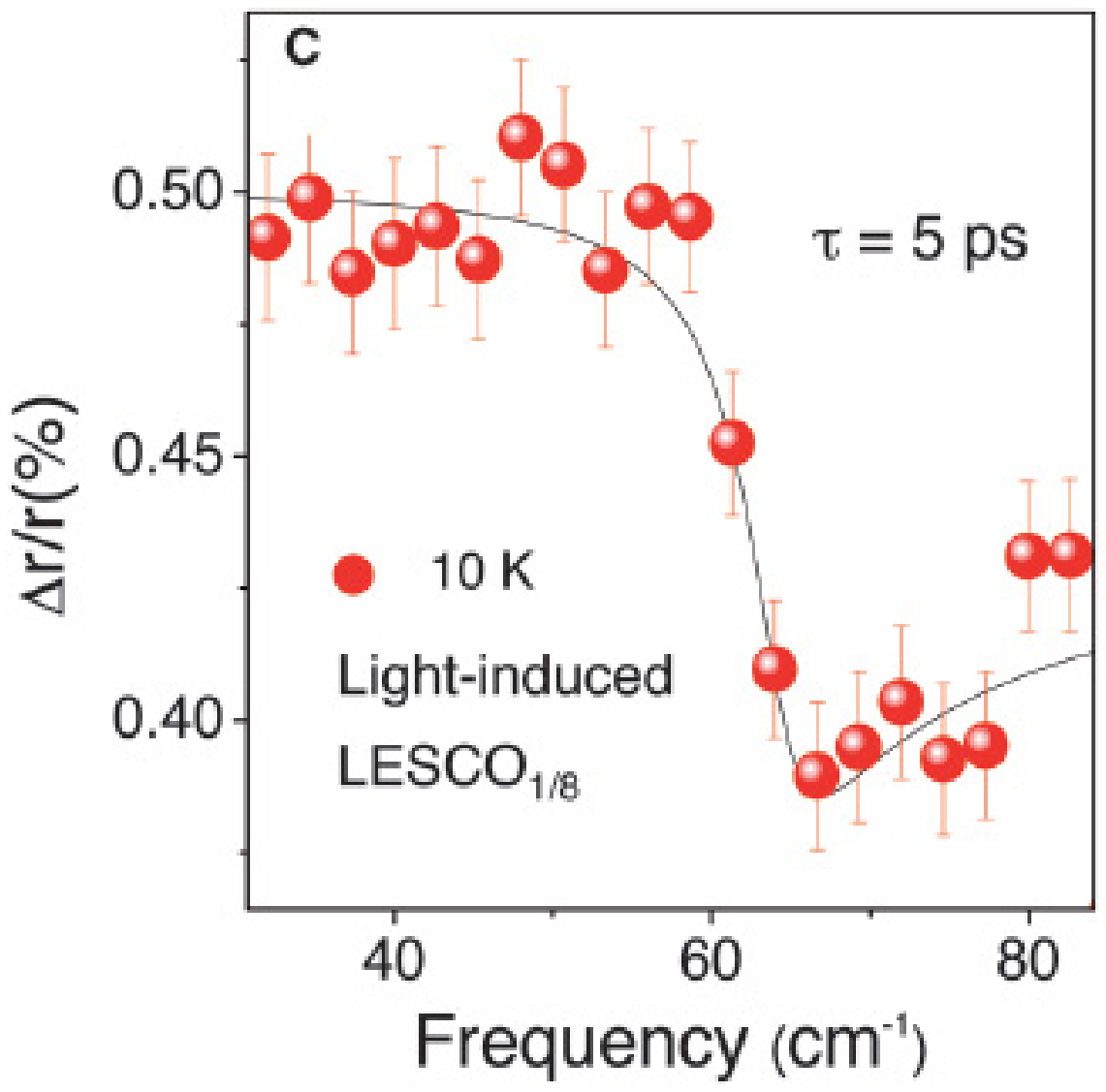}
\includegraphics[width = 0.49\columnwidth]{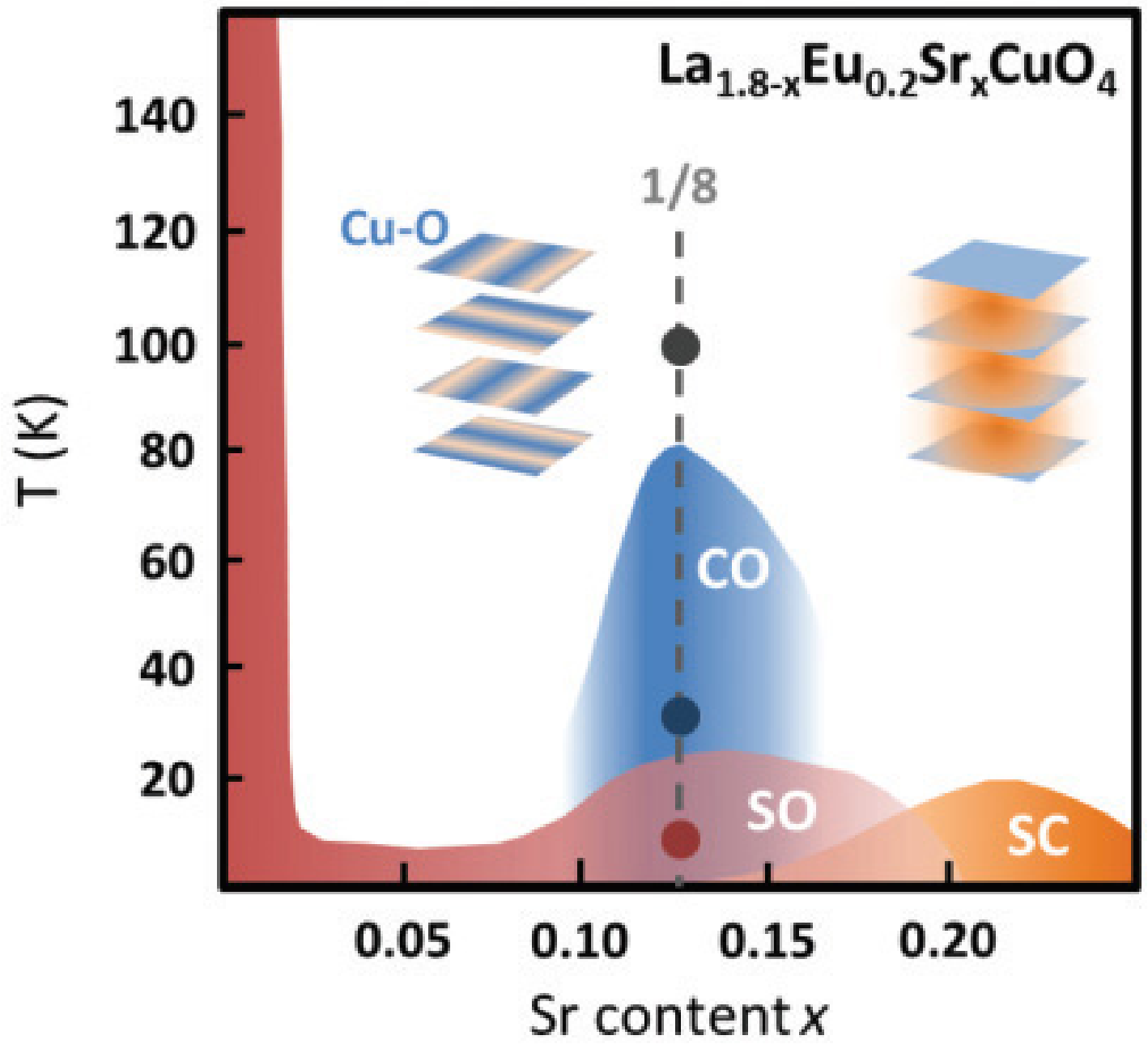}
\caption{Left panel, taken from Ref.~\cite{Fausti-2011} with the original caption: Transient $c$-axis reflectance of LESCO$_{1/8}$, normalized
to the static reflectance. Measurements are taken at 10 K, after excitation with IR pulses at 16 $\mu$m
wavelength. The appearance of a Josephson plasma edge at 60 cm$^{-1}$ demonstrates that the photoinduced state
is superconducting. Right panel, taken from Ref.~\cite{Hunt-2015}  with the original caption: Phase diagram of LESCO, based
on Supplemental Material at http://link.aps.org/supplemental/
10.1103/PhysRevB.91.020505,  indicating regions of bulk superconductivity (SC) and
static spin (SO) and charge (CO) order. The static stripes suppress
$c$-axis coupling of the CuO$_2$ planes (inset cartoon, left), with bulk
superconductivity restored at dopings in which the stripe order
is reduced (inset cartoon, right).}
\label{figA}
\end{figure}

It is impossible to do justice here to the complete literature relevant to these
experiments, which is vast because the interaction of spin-ordering,
charge-ordering, and superconductivity has been one of the
most central issues in condensed matter physics for more than 30 years. There 
is reason to believe, in fact, that spin- and charge-ordering in 
stripes is closely related to the origin of high-temperature superconductivity. We
will instead focus on just the papers that are most directly relevant to
light-induced superconductivity in the specific material La$_{1.675}$Eu$%
_{0.2}$Sr$_{0.125}$CuO$_{4}$ (LESCO$_{1/8}$)~\cite
{Fausti-2011,Forst-2014,Hunt-2015,Rajasekaran-2018,Cavalleri-2018}, represented by the results of Refs.~\cite{Fausti-2011} and \cite{Hunt-2015} shown in Fig.~\ref{figA}.

The work of Refs.~\cite{Fausti-2011,Forst-2014,Hunt-2015,Rajasekaran-2018,Cavalleri-2018}, and that 
cited in these papers, indicate that coherent 3-dimensional
light-induced superconductivity emerges when the competing coherent 3-dimensional phase of
in-plane stripes is ``melted'' by an ultrafast laser pulse. 

Here we will
consider a simple time-dependent Ginzburg-Landau model of these competing
phases: 
\begin{eqnarray}
-\tau _{1}\frac{dn_{1}}{dt} =\frac{\partial F}{\partial n_{1}}n_{1}\quad
&,& \quad -\tau _{2}\frac{dn_{2}}{dt}=\frac{\partial F}{\partial n_{2}}n_{2} 
\end{eqnarray}
\begin{eqnarray}
F =-a_{1} n_{1}+\frac{1}{2}b_{1}n_{1}^{2}+q_{1}^{2}A(t)^{2}n_{1} 
-a_{2} n_{2}+\frac{1}{2}b_{2}n_{2}^{2}+q_{2}^{2}A(t)^{2}n_{2}+c \, n_{1}n_{2}
\end{eqnarray}
where $n_{1}$ and $n_{2}$ respectively represent condensate densities for the
3-dimensional superconducting and stripe phases. 

All the coefficients $a_i$, $b_i$, $q_i$, and $c$ are in principle temperature as 
well as materials dependent (with $q_i$ also frequency dependent). The terms involving $a_i$ and $b_i$ are standard 
in a Ginzburg-Landau description of superconductors (and various other systems). 
The terms involving $q_{i}^{2}A(t)^{2}n_{i}$ result from a Ginzburg-Landau 
description averaged over one wavelength of the laser radiation with
\begin{eqnarray}
\psi^{*} \frac{1}{2m} \left( -i \mathbf{\nabla}- \frac{q_{eff}}{c} \mathbf{A}(t) \right) ^2 \psi \longrightarrow q^{2}A(t)^2 n \quad , \quad n= \psi^{*} \psi
\end{eqnarray}
if the wavelength of the radiation is large compared to the length scale for variations in the order parameter. 
(The bare kinetic energy from $\mathbf{\nabla}^2$ is contained in the other parameters, with any shift in kinetic energy approximately 
absorbed in the $q_{i}^{2}A(t)^{2}n_{i}$ term. For a laser field oscillating with a single frequency $\omega$, 
the average intensity is proportional to $A(t)^{2}$.) 
We note that (i) charge- and spin-density waves are similar in some respects to superconductivity, so the symmetry
in $F$ is natural for a simplest model in the present context, and (ii) the essential point is just that both the stripe and 
superconducting phases couple to an oscillating electromagnetic field (with intensity 
proportional to $A^2$). 
The term $c \, n_{1}n_{2}$ describes the fact that two competing phases -- with very different length scales, textures, 
and even topologies -- must both recruit the same electrons, so that one tends to frustrate the other, as has long been known.
The form for the time dependence is chosen 
because it gives an exponentially fast rise time for small $n_i$, and also an exponentially slow decay time, so that $n_i$ remains positive.
An extra feature of the model is that small random fluctuations are introduced in each $n_i$ at each time step, to simulate the physical 
(thermal and quantum) fluctuations of an order parameter. Without these fluctuations, $n_i$ could never recover after going to zero. 
Finally, we note that, with this form for the time dependence, there is an exponentially fast approach to equilibrium for both phases, 
from either below or above the equilibrium values of $n_i$, 

The time-dependent equations are 
\begin{eqnarray}
\tau _{1}\frac{dn_{1}}{dt} = \left[ a_{1} -\left(
b_{1}n_{1}+q_{1}^{2}A(t)^{2}+c \, n_{2}\right) \right] n_{1} \\
\tau _{2}\frac{dn_{2}}{dt} =\left[ a_{2} -\left(
b_{2}n_{2}+q_{2}^{2}A(t)^{2}+c \, n_{1}\right) \right] n_{2} 
\end{eqnarray}
with the following time-independent solutions: Either $n_{1}=0$ or  
\begin{eqnarray}
n_{1}=\frac{a_{1} -q_{1}^{2}A(t)^{2}-c \, n_{2}}{b_{1}}
\;
\end{eqnarray}
and either $n_{2}=0$ or else
\begin{eqnarray}
n_{2}=\frac{a_{2} -q_{2}^{2}A(t)^{2}-c \, n_{1}}{b_{2}}
\;
\end{eqnarray}
(with unphysical negative solutions excluded and never reached in a numerical solution). 

One can solve for $n_{1}$
and $n_{2}$, but what is most interesting here is the qualitative behavior:

(1) If  $q_{2}^{2}A^{2}>a_{2}$, there will be a
nonthermal ``melting'' of an initial stripe phase. Then if $a_{1}>q_{1}^{2}A^{2}$, the superconducting phase
will emerge, as observed.

(2) Depending on the specific parameters for a given material and set of conditions, 
there may be no ordered phase, or either, or both coexisting, as is consistent with 
a large body of experimental and theoretical work.

(3) There is a reciprocity inherent in the free energy: The superconducting
phase is just as effective in blocking the stripe phase as vice-versa, in
the sense that the same coefficient $c$ is involved. This can
explain why the superconducting phase persists for an extremely long time in
the low-temperature results of Refs.~ \cite{Fausti-2011} and  \cite{Hunt-2015} -- at least 100 picoseconds
and perhaps up to nanoseconds and longer, for temperatures below about 25 K. 

(4) However, the coefficient $c$ depends on the character of both phases. This appears to be reflected in the experimental results above about 25 K~\cite{Hunt-2015}, 
where the spin- and charge-ordering undergoes a change of character to a different phase, as can be seen in the right-hand panel of Fig.~\ref{figA}, taken from Ref.~\cite{Hunt-2015}. 
According to Ref.~\cite{Hunt-2015}, ``Below T$_{\mathrm{SO}}$ , the lifetimes remain temperature independent.
Above T$_{\mathrm{SO}}$, where only static charge order remains, the lifetime drops exponentially with base temperature.... The exponential dependence of the relaxation between
T$_{\mathrm{SO}} < T < T_{\mathrm{CO}} $ can be reconciled with the expected kinetic
behavior for a transition between two distinct thermodynamic
phases separated by a free energy barrier.'' If $c$ is smaller for the higher-temperature phase, then the metastability of the superconducting phase will be weakened, 
permitting a relatively rapid activated transition back to the more stable phase.

\begin{figure}[tbp]
\centering
\includegraphics[width = \columnwidth]{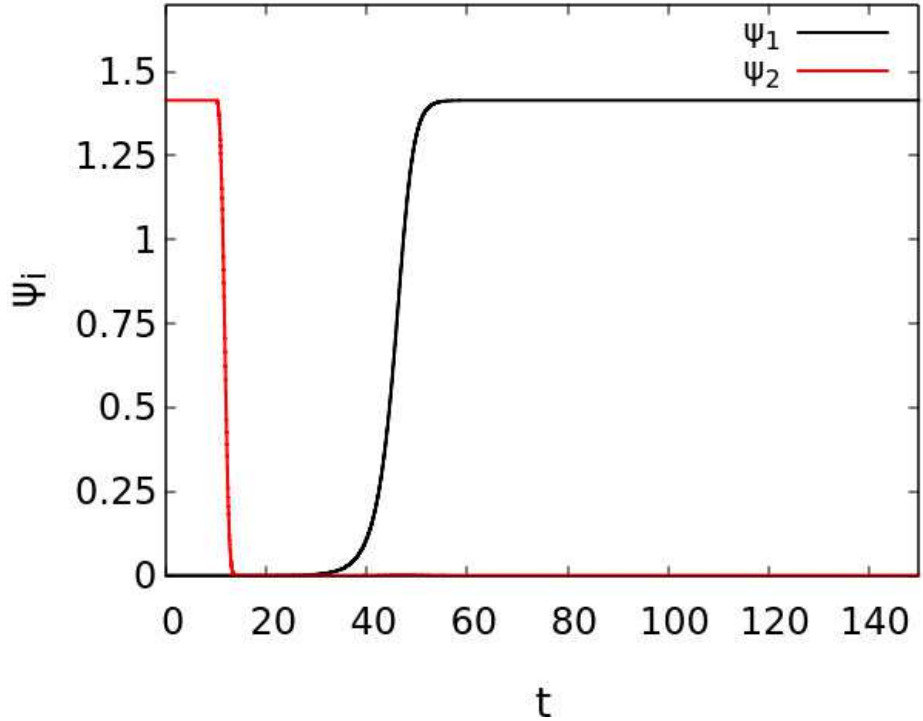}
\caption{Simulation for two competing phases with order parameters $\psi_1$ and $\psi_2$. 
The order parameter $\psi_i$, which for the present simple model is real, is related to the condensate density by $n_i = \psi_i^2$.
When the dominant phase $2$ (``stripes'') is suppressed by the
laser pulse between $t=10$ and $t=30$, the other phase $1$ (``superconductivity'') emerges and persists
indefinitely after the laser pulse has finished.}
\label{fig1}
\end{figure}

A typical numerical solution of the above equations for a qualitative model is shown in Fig.~\ref{fig1}, with a model laser pulse having the form
\begin{eqnarray}
A(t) = A_{0} \sin \left( \pi (t-t_0) / 2 \tau \right) \sin \left( \omega (t-t_0 ) \right) \quad , \quad t_0 < t < t_0 + \tau
\end{eqnarray}
where $t_0 = 10$, $\tau = 20$, $A_0 = 10$, and $\omega = 2$. (This form closely resembles a Gaussian envelope modulated by oscillations with frequency $\omega$.) 
The dominant phase
(``stripes'') has parameters $\tau_2 = 5$, $a_2 = 2$, $b_2=1$, $q_2^2=2$, and the other phase (``superconductivity'') has $\tau_1=5$, $a_1 =1.8$, $b_1=0.9$, $q_1^2=0$, with $c=2$. 

Both of the main qualitative features of Fig.~\ref{fig1}  are similar to what
is observed in the experiments: First, when the dominant phase is suppressed
by the laser pulse, the other phase quickly emerges. Second, the other phase
persists for an indefinite period of time after the pulse is finished, in a robust metastable state. 

The present model can clearly be extended in many ways, with realistic models constructed for specific materials, 
but the present note is meant only to demonstrate its qualitative potential.

\section{Acknowledgements}

We have benefitted from many discussions with Ayman Abdullah-Smoot, Michelle
Gohlke, David Lujan, and James Sharp.

\bigskip \noindent
Author contribution statement: Roland E. Allen originated the project in consultation with M. Ross Tagaras and Jian Weng, who performed the calculations.


\begin{thebibliography}{9}
\bibitem{Fausti-2011} D. Fausti, R. I. Tobey, N. Dean, S. Kaiser, A. Dienst, M.
C. Hoffmann, S. Pyon, T. Takayama, H. Takagi, A. Cavalleri, ``Light-Induced
Superconductivity in a Stripe-Ordered Cuprate'', Science 331, 189 (2011).

\bibitem{Forst-2014} M. F\"{o}rst, R. I. Tobey, H. Bromberger, S. B.
Wilkins, V. Khanna, A. D. Caviglia, Y.-D. Chuang, W. S. Lee, W. F.
Schlotter, J. J. Turner, M. P. Minitti, O. Krupin, Z. J. Xu, J. S. Wen, G.
D. Gu, S. S. Dhesi, A. Cavalleri, and J. P. Hill, ``Melting of Charge
Stripes in Vibrationally Driven La$_{1.875}$Ba$_{0.125}$CuO$_{4}$: Assessing
the Respective Roles of Electronic and Lattice Order in Frustrated
Superconductors'', Phys. Rev. Lett. 112, 157002 (2014).

\bibitem{Hunt-2015} C. R. Hunt, D. Nicoletti, S. Kaiser, T. Takayama, H.
Takagi, and A. Cavalleri, ``Two distinct kinetic regimes for the relaxation
of light-induced superconductivity in La$_{1.675}$Eu$_{0.2}$Sr$_{0.125}$CuO$%
_4$'', Phys. Rev. B 91, 020505(R) (2015).

\bibitem{Rajasekaran-2018} S. Rajasekaran, J. Okamoto, L. Mathey, M.
Fechner, V. Thampy, G. D. Gu, A. Cavalleri, ``Probing optically silent
superfluid stripes in cuprates'', Science 359, 575 (2018).

\bibitem{Cavalleri-2018} Andrea Cavalleri, ``Photo-induced
superconductivity'', Contemporary Physics 59, 31 (2018).
\end{thebibliography}
\end{document}